\documentclass[sigplan,9pt,screen,natbib,nonacm]{acmart}
\pdfoutput=1    
\setcopyright{none}

\usepackage{hyphenat}

\usepackage{mdwtab}
\usepackage{mdwlist}
\usepackage{mathenv}
\usepackage{amsmath}
\usepackage{comment}
\usepackage{url}
\usepackage{prooftree}

\usepackage{listings}

\lstdefinestyle{sOcaml}{language=[Objective]Caml,
  morekeywords={effect,perform,locus},
  literate={+}{{$+$}}1 {/}{{$/$}}1 
           {=}{{$=$}}1
           {>}{{$>$}}1 {<}{{$<$}}1
           {<>}{$\not=$}1
           {->}{{$\rightarrow$}}2 {>=}{{$\geq$}}2 {<-}{{$\leftarrow$}}2
           {<=}{{$\leq$}}2
           {==>}{{$\mapsto$}}2
           {|}{{$\mid$}}1
           {'a}{$\alpha$}1
           {'b}{$\beta$}1
           {'c}{$\gamma$}1
           {'w}{$\omega$}1
           {'r}{$\rho$}1
           {'state}{$\sigma$}1
           {fn}{$\lambda$}1
           {:=}{\ensuremath{\mathrel{{:}{=}}}}2
           {...}{\ldots}2
           {empty}{\ensuremath\varnothing}1
           {\#\#+}{\color{red}}1
           {\#\#-}{\color{black}}1
           {\#\#\#}{{$\leadsto$}}3
}
\newcommand\coleqeq{\mathrel{\mbox{::=}}}
\newcommand\I{\ensuremath{\mathord{\mathsf{int}}}}
\newcommand\B{\ensuremath{\mathord{\mathsf{bool}}}}
\renewcommand\U{\ensuremath{\mathord{\mathsf{unit}}}}
\newcommand{\tref}[1]{#1\,\mathsf{ref}}
\newcommand{\tptr}[1]{#1\,\mathsf{ptr}}
\newcommand{\tconst}[1]{#1\,\mathsf{const}}
\newcommand{\eref}{\mathord{\mathsf{ref}}\,}
\newcommand{\be}[1]{\mathsf{let}\:#1\:\mathsf{in}\:}
\newcommand{\Tr}[1]{\ensuremath{\lceil #1 \rceil}}
\newcommand{\TC}[1]{\ensuremath{\lfloor #1 \rfloor}}
\newcommand{\Den}[1]{\ensuremath{[\![\, #1\, ]\!]}}

\lstnewenvironment{code}{\lstset{basicstyle={\linespread{1.01}\sffamily\normalsize}}}{}

\lstMakeShortInline[columns=fullflexible]|

\newcommand{\aside}[1]{\ignorespaces}

\begin{document}
\title{Do Mutable Variables Have Reference Types?}

\author{Oleg Kiselyov}
\orcid{0000-0002-2570-2186}
\affiliation{%
  \institution{Tohoku University}
  \country{Japan}}
\email{oleg@okmij.org}

\begin{abstract}
Implicit heterogeneous metaprogramming (a.k.a. \emph{offshoring}) is
an attractive approach for generating C with some correctness
guarantees: generate OCaml code, where the correctness guarantees are
easier to establish, and then map that code to C. The key idea is that
simple imperative OCaml code looks like a non-standard notation for C.
Regretfully, it is false, when it comes to mutable variables.  In the
past, the approach was salvaged by imposing strong ad hoc
restrictions. The present paper for the first time investigates the
problem systematically and discovers general solutions needing no
restrictions. In the process we explicate the subtleties of modeling
mutable variables by values of reference types and arrive at an
intuitively and formally clear correspondence. We also explain C
assignment without resorting to L-values.
\end{abstract}
\maketitle

\section{Introduction}
\label{s:intro}


Generating C (or other such low-level language) is inevitable if we
want the convenience and guarantees of abstractions~-- and we want the
code that runs in a constrained environment (e.g., a low-powered robot);
involves OpenMP, OpenCL (i.e., GPGPU) or AVX512 instructions;
benefits from profitable but highly domain-specific optimizations
typical in HPC. In fact, we have done all of the above, using the
offshoring technique first proposed in \cite{eckhardt-implicitly} and
re-thought and re-implemented in \cite{generating-C}. The key idea of
offshoring, explained below, is the close correspondence between
imperative OCaml and C code.

Mutable variables of C is the biggest stumbling block: the
straightforward mapping of OCaml variables of reference types to C
mutable variables is insidiously wrong, when it comes to aliasing. In
the past, pitfalls were steered around of by imposing strong, ad hoc
restrictions~-- which made generating C code with
mutable variables of pointer types, for example, out of reach.

We propose a better mapping between reference-type and 
mutable variables that needs no
restrictions and hence widens the scope of offshoring.

After introducing offshoring, the paper explains the problem of
generating code with mutable variables, and, in
\S\ref{s:current-offshoring}, its current, imperfect resolution.
\S\ref{s:no-mut-var} and \S\ref{s:final-translation} each introduce
new proposals, improving the state of the
art. \S\ref{s:final-translation}'s approach is
the most general, intuitive, easier to show correct, and insightful.
It applies to any other language which uses values of reference types
to model mutable variables, such as F\# and SML.

\section{Offshoring}

Offshoring turns homogeneous metaprogramming~-- generating OCaml in
OCaml~-- into heterogeneous: generating C. The key idea is that simple
imperative subset of OCaml may be regarded as a different notation for
C. Taking the running example of vector addition from
\cite{generating-C}, contrast the OCaml code
\newpage
\begin{code}
let addv = fun n vout v1 v2 ->
   for i=0 to n-1 do 
       vout.(i) <- v1.(i) + v2.(i) done
\end{code}
and the C code:
\begin{code}
void addv(int n, int* vout, int* v1, int* v2) {
 for(int i=0; i<=n-1; i++)
    vout[i] = v1[i] + v2[i];
}
\end{code}
The similarity is so striking that one may argue that OCaml's |addv| is
C's |addv|, written in a different but easily relatable way.
Offshoring is the facility that realizes such correspondence between a
subset of OCaml and C (or other low-level language).  With offshoring,
by generating OCaml we, in effect, generate C. 

With MetaOCaml, we may statically ensure the generated OCaml code
compiles without errors. If we can map OCaml to C while preserving the
guarantees, we in effect obtain the assured C code generator. Needless
to say, the mapping ought to preserve the dynamic
semantics/behavior.

It should be stressed that the mapping from OCaml to C in not
total. We are not aiming to translate all of OCaml to C~-- only a
small imperative subset. That is why we are not concerned with
closures, recursion, user-defined datatypes, let
alone more complicated features. Therefore, we generate
efficient C that does not need any special run-time. We are
not aiming to generate every C feature either. After all, like other
languages C is redundant: many differently phrased
expressions compile to the same machine code.  The supported
subset of OCaml and C, if small, should still be useful~-- and it
proved to be in our experience, for generating HPC and embedded code.

\section{Problem with Mutable Variables}

However small the mappable OCaml subset may be, its range should
include mutable variables, which are pervasive in C. Offshoring would
hardly be useful otherwise. Thus the central problem is what should be
the OCaml code that maps to C code with mutable variables. OCaml
values of reference types are not a straightforward match of C mutable
variables~-- as one realizes upon close inspection.  Hereby we
undertake the systematic investigation of the problem.

\subsection{Formalization}
\label{s:formalization}

We introduce the calculus ICaml to delineate the minimal relevant
first-order imperative
subset of OCaml. (The subset of OCaml used in
offshoring \cite{generating-C} is not much bigger, adding loops and
conditionals.)

\begin{figure}[htbp]
\begin{tabular}[Ct]{lMl}
Variables & x,y,z \\
Base Types & b \coleqeq \I \mid \B
\\
Types & t \coleqeq \U \mid b \mid \tref{t}
\\
Constant Types &
s \coleqeq t \mid t\to s
\\
Expressions &
e \coleqeq x \mid c_0 \mid c_1\, e \mid c_2\,e\ e \mid e; e \mid \be{x = e}e
\end{tabular}
\caption{Calculus ICaml. For constants $c_i$ see
  Fig. \ref{f:IC-constants}.}
\label{f:ICaml}
\end{figure}

Most of it is self-explanatory. Constants $c_i$ have to be applied to
$i$ arguments to be considered expressions. We use the customary infix
notation for such applications where appropriate. 

The calculus is Church-style: all (sub)expressions are annotated with
their types. To avoid clutter however, we mostly elide types, where
they can be easily understood. The type system is entirely standard
and elided to save space. The dynamic semantics is also standard.

\begin{figure}
\begin{eqnarray*}[cl:cl]
1,2,3,\ldots   &: \I
&
\eref &: t\to \tref{t}
\\
true, false &: \B
&
! &: \tref{t}\to t
\\
+ &: \I\to\I\to\I
&
:= &: \tref{t}\to t\to \U
\end{eqnarray*}
\caption{The constants $c_i$ of ICaml and their types. Their arity $i$
  is the number of arrows in their type. For instance, $!$ is a
  1-arity constant $c_1$, and $:=$ is $c_2$. Only constants have arrow
  types.  We may silently add other similar constants.}
\label{f:IC-constants}
\end{figure}

The calculus CoreC, Fig.~\ref{f:CoreC}, models the relevant subset of
C: the target of the offshoring mapping. It is also entirely
standard. The static and dynamic semantics of ICaml and CoreC are shown
in full (in tagless-final style) in the file \url{refcalculi.ml}
accompanying the paper
(see also \url{http://okmij.org/ftp/tagless-final/refcalculi.ml}).

\begin{figure}[htbp]
\begin{tabular}[Ct]{lMl}
Variables & x,y,z \\
Base Types & b \coleqeq \I \mid \B
\\
Types & t \coleqeq \U \mid b
\\
Constant Types &
s \coleqeq t \mid t\to s
\\
Expressions &
e \coleqeq x \mid c_0 \mid c_1\, e \mid c_2\,e\ e \mid e; e \\
& \qquad t\; x = e; e \mid x := e
\end{tabular}
\caption{Calculus CoreC. Its constants $c_0$ and the 2-arity constant $+$
  are the same as those in Fig. \ref{f:IC-constants}.}
\label{f:CoreC}
\end{figure}

The calculus permits expressions like $(\I\,x=1+2; x+3)+4$, which is
invalid C. However, with the simple post-processing step of lifting
variable declarations (always possible if the names are unique, which
is assumed), it becomes 
|int x; (x=1+2,x+3)+4|, which is proper C.

The calculi ICaml and CoreC are quite alike; the main difference is in
mutation. To emphasize the distinction, we show the (natural
deduction) typing derivation of integer assignment in the two calculi.

\medskip
\noindent
\begin{prooftree}
x: \tref\I
\quad
e: \I
\quad
:=\:: \tref{\I}\to\I\to\U
\justifies
x := e\:: \U
\end{prooftree}
\quad
\begin{prooftree}
x: \I
\quad
e: \I
\justifies
x := e\:: \U
\end{prooftree}

\medskip
\noindent
We see in particular that whereas $x := x$ in ICaml cannot be typed,
this expression is well-typed and meaningful (as a no-op) in
CoreC. Assignment in CoreC is a special form; therefore, the same $x$
on different sides may have different meanings~-- traditionally
described by the terms L-value and R-value.

The calculi ICaml and CoreC resemble the corresponding calculi in
\cite{eckhardt-implicitly}, but are much, much simpler~-- and free from
the severe restriction on initializers being constants.

\subsection{Extant Offshoring Translation}
\label{s:current-offshoring}

We now state the translation $\Tr{\cdot}$ from a (type-annotated)
expression of ICaml to a type-annotated expression of CoreC, 
Fig.~\ref{f:offshoring-naive},
\begin{figure}
\begin{eqnarray*}[c@{\;=\hskip 1em}c]
\Tr{x:t} & x:\Tr{t}\\
\Tr{! (x:t)} & x:\Tr{t}\\
\Tr{(x:t) := e} & x:\Tr{t} := \Tr{e}\\
\Tr{\be{x:t = \eref e}e'} & \Tr{t}\; x = \Tr{e}; \Tr{e'}\\
\Tr{\be{x:t = e}e'} & \Tr{t}\; x = \Tr{e}; \Tr{e'}
\end{eqnarray*}
\caption{Naive offshoring translation}
\label{f:offshoring-naive}
\end{figure}
with the rest being homomorphism. Here $\Tr{\tref{t}}=t$ and
$\Tr{t}=t$ otherwise. This is basically the translation
proposed in \cite{eckhardt-implicitly}, adjusted for (many)
differences in notation. As an example, the ICaml expression
\[
\be{x=\eref 0}x:={!x}+1
\]
is translated to
\[
\I\, x=0; x:=x+1
\]

The translation clearly expresses the idea that an OCaml value of
a reference type bound to a variable is a model of mutable variables in
C. It is also clear that the translation is partial: ICaml code like
$!(\eref 0)$ or $(\eref 0) := 1$ is not translatable. The translation is also
non-compositional: variable references are translated differently if
they appear as the first argument of $!$ or the assignment
operation. If we add constants with arguments
of reference types, like \textsf{incr}, the translation has to be amended.

There is also a subtle and serious problem with the translation as
written. Applying it to
\begin{eqnarray}
\label{e:aliasing}
\be{x=\eref 0}\be{y=x}y:=41;{!x}+1
\end{eqnarray}
would give
\[
\I\, x=0; \I\, y=x; y:=41; x+1
\]
which has a different meaning. Whereas
the ICaml expression evaluates to 42, its CoreC translation
returns 1.

The root of the problem is the difference in meaning between 
$\be{y=x}\ldots$ in ICaml and $t\; y=x;$ in CoreC. In the latter case,
a new mutable variable is allocated whose initial contents is the current
value of $x$. Then $y$ and $x$ are mutated independently. On the other
hand, $\be{y=x}\ldots$ allocates no new reference cell; it mere
introduces a new name, $y$, for the existing reference cell named $x$.
One may informally say that in ICaml, names of mutable cells are
first-class. 

Although the interpretation of names in ICaml and CoreC differs, as we
have just seen, the difference fades in restricted contexts. The
offshoring translation can be made meaning-preserving, after all~-- if
we impose restrictions that preclude aliasing. The paper
\cite{eckhardt-implicitly} never mentions that fact
explicitly. However, if we carefully examine the typing rules in its
Appendix A2, we discover the silently imposed restrictions: only
base-type references, and only base-type let-bindings (sans the
dedicated expression $\be{x=\eref e}e'$ for creating references). The
fixed offshoring translation is shown in Fig.~\ref{f:offshoring-extant}.

\begin{figure}
\begin{eqnarray*}[c@{\;=\hskip 1em}c]
\Tr{x:b} & x:b\\
\Tr{! (x:\tref{b})} & x:b\\
\Tr{(x:\tref{b}) := e} & x:b := \Tr{e}\\
\Tr{\be{x:\tref{b} = \eref e}e'} & b\; x = \Tr{e}; \Tr{e'}\\
\Tr{\be{x:b = e}e'} & b\; x = \Tr{e}; \Tr{e'}
\end{eqnarray*}
\caption{The extant offshoring translation, with clearly stated restrictions}
\label{f:offshoring-extant}
\end{figure}

This is the (core of the) offshoring translation used in the current
MetaOCaml: BER N111. It has been used in all offshoring applications
so far, many of which are mentioned in \cite{generating-C}, which also
details simpler use cases.

Although the translation proved more or less adequate for numeric code, it
is clearly severely restrictive: it is impossible, for example, to
generate C code with pointer-type function 
arguments or pointer-type mutable variables~-- or even
represent pointer types to start with. Occasionally we had
to fiddle with a generator to make it produce the offshorable
OCaml code.
Also, the translation
of the problematic
\eqref{e:aliasing} is not fixed: merely outlawed.

Each of the two following sections propose a new translation,
overcoming the drawbacks of the state of the art.

\section{C without Mutable Variables?}
\label{s:no-mut-var}

Paper \cite{generating-C} briefly mentions, merely on two examples and
without details or formalization, an alternative translation: avoiding
mutable variables altogether. In terms of the present paper, the
calculus CoreC becomes unnecessary: ICaml as is gets mapped to the
surface syntax of C. We now present the translation formally and systematically,
noting its advantages and disadvantages. The disadvantages motivate
the proposal in \S\ref{s:final-translation}.

The key idea is that C already has the analogue of ICaml values of
reference type: arrays. Expressions of reference types of ICaml are C
pointer expressions.
Assuming the lifting transformation mentioned in
\S\ref{s:formalization}, the mapping 
$\TC{\cdot}$ from ICaml to C is as follows:

\begin{eqnarray*}[c@{\;=\hskip 1em}Tc]
\TC{\eref (e:t)} & |t z[1] = {$\TC{e}$}; z|\qquad   (\textsf{z} is fresh)\\
\TC{! e} &         |*$\TC{e}$|\\
\TC{e := e'}   &   |*$\TC{e}$ := $\TC{e'}$|\\
\TC{\be{x:\tref{t} = e}e'} & |t * const x = $\TC{e}$; $\TC{e'}$|\\
\TC{\be{x:b = e}e'} & |b x = $\TC{e}$; $\TC{e'}$|
\end{eqnarray*}
One is reminded of Algol68, in which \textsf{.int x := 1} is the abbreviation
for \textsf{.ref.int x = .local.int := 1}, where |.local.int| is the stack
allocator of an integer (similar to \textsf{alloca} in C).

The earlier example
\[
\be{x=\eref 0}x:={!x}+1
\]
now looks like
\[
\texttt{int * const x = (int z[1]={0}; z);\; *x = *x + 1;}
\]
or, after variable lifting
\[
\texttt{int z[1] = {0};\; int * const x = z;\; *x = *x + 1;}
\]
The problematic \eqref{e:aliasing}, that is,
\[
 \be{x=\eref 0}\be{y=x}y:=41;{!x}+1
\]
becomes
\begin{tabular}[C]{l}
\tt
int z[1] = {0}; int * const x = z;\\
\tt
int * const y = x; *y = 41; *x + 1
\end{tabular}
and returns the same result as the ICaml code. 

There are no longer any restrictions to base types; adding 
constants like |incr| is easy. Arrays, conditionals, loops are
straightforward as well.

Other extensions are more problematic, however. If we extend ICaml
with composite data structures (e.g., to express linked lists),
functions returning values of reference types, or global variables
or arguments of higher-reference types~-- we have to worry about the
lifetime of reference cells allocated by $\eref e$. (These extensions
are not common in numeric computing however.) They have to be
allocated on heap, and managed somehow (e.g., via reference counting).
We have to stress, however, that resource/memory management is the
problem that has to be dealt with at higher levels of abstractions; by
the time of offshoring, the code should already be assured
resource-safe. See \cite{session-types}
for an example of such resource-safety assurances.


The translation results in highly unidiomatic C code, which, from
personal experience, provokes negative reaction and strong doubts
about correctness. A reviewer suggested a slight adjustment, which
makes the result look a bit more familiar.
\begin{eqnarray*}[c@{\;=\hskip 1em}Tc]
\TC{\eref (e:t)} & |alloca($\TC{e}$)|
\end{eqnarray*}
where |alloca| could be avoided by allocating a fresh variable in
scope. Thus the running example
\[
\be{x=\eref 0}x:={!x}+1
\]
becomes
\[
\texttt{int z = 0;  int * const x = \&z;  *x = *x + 1;}
\]
which looks a bit more like conventional C.

One of the advantages of the present translation is that
\[
\TC{\tref{t}} = \tconst{\TC{t} *} 
\]
That is, reference types of ICaml map directly to pointer types in C.
On the downside, we do not represent mutable variables of C or CoreC
directly in ICaml. The disadvantage has a practical side: as seen from
the translation examples, each reference-type variable of ICaml is
translated to \emph{two} variables in C: one holds the content and is
mutated, and the other is the pointer to the former. The C code hence
needs twice as many variables~-- and twice as much stack storage for
them. In simple code, a C compiler can notice variables that are not
mutated and effectively inline them, removing the need to store
them. However, we are aiming to generate very complicated code. Take,
for example matrix-matrix multiplication from our past work: applying
standard techniques to make it fast results in thousands of lines of C
code. There, C compiler may not see that some variables are redundant.

Can mutable variables in (Core)C be represented directly in ICaml? Can
the translation hence map those ICaml variables directly, one-to-one,
to CoreC mutable variables, without allocating pointers to them?  Can
we intuitively and formally be confident in the translation, even for
arbitrarily complex reference types?  The following section shows the
answer.

\section{Mutable Variables and Reference Types}
\label{s:final-translation}

We have just seen the translation from ICaml to C that maps ICaml
variables of reference types to constant-pointer--type variables of C.
We now present the translation that relates reference type variables
of ICaml and mutable variables of C. It requires no restrictions,
produces idiomatic C code, and gives insight into the nature of
mutable variables.

An easy way to obtain a translation with mutable variables in its
range is to start with the straightforward inverse mapping from CoreC
to ICaml. Unfortunately, it is very much not surjective (and if we
extend CoreC with pointer types, it becomes non-injective). Therefore,
inverting it is problematic. Still the CoreC to ICaml mapping gives a
hint. Other hints come from looking at the denotational semantics
(tagless-final interpreters) of ICaml and CoreC: the file
\url{refcalculi.ml} in the accompanying code mentioned earlier.  We
notice that $\be{x=\eref e}e'$ has exactly the same denotation (as the
function of the denotations of $e$ and $e'$) as $t\; x=e; e'$ in
CoreC. Therefore, if the mutable variable $x$ introduced by $t\; x=e;
e'$ is not actually mutated in $e'$, it has the meaning of the
let-binding in ICaml.

To formulate the new translation, we extend
CoreC with pointer types and corresponding operations, obtaining the
calculus CoreCE.
\begin{figure}
\begin{tabular}[Ct]{lMl}
Variables & x,y,z \\
Base Types & b \coleqeq \I \mid \B
\\
Types & t \coleqeq \U \mid b \mid \tptr{t}
\\
Binder Types & u \coleqeq t \mid \tconst{t}
\\
Constant Types &
s \coleqeq t \mid t\to s
\\
Expressions &
e \coleqeq x \mid c_0 \mid c_1\, e \mid c_2\,e\ e \mid e; e \\
& \qquad u\; x = e; e \mid \&x
\end{tabular}

\medskip
Additional Constants
\begin{eqnarray*}[cl:cl]
* &: \tptr{t}\to t
&
\leftarrow &: \tptr{t}\to t\to \U
\end{eqnarray*}
\caption{Calculus CoreCE. Constants $c_0$ and $+$ are 
  same as those in Fig. \ref{f:IC-constants}.}
\label{f:CoreCE}
\end{figure}
Assignment is no longer a special expression form: it is a
function application. Its both arguments are ordinary function
application arguments, with no need to introduce L-values. Other
pointer-taking functions like incr can be added at will. We also add
constant (binder) types, to indicate that some variables are
immutable. In surface C,
$e\leftarrow e'$ is to be rendered as |*e = e'|~-- which is a key to
understanding C assignment without resorting to L-values. Also, in surface C we
abbreviate |*&x| to just |x|. (We may also leave |*&x| as is: it is
valid.) 
The type system and dynamic semantics are fairly standard: see
\url{refcalculi.ml}.

The offshoring translation $\Tr{\cdot}_L$ is parameterized by the set
$L$ of mutable variables in scope:
\begin{eqnarray*}[c@{\;=\hskip 1em}c]
\Tr{x:t}_L & x:\Tr{t}\quad x\not\in L\\
\Tr{x:t}_L & \&x:\Tr{t}\quad x\in L\\
\Tr{!} & *\\
\Tr{:=} & \leftarrow\\
\Tr{\be{x:\tref{t} = \eref e}e'}_L & 
\Tr{t}\; x = \Tr{e}_L; \Tr{e'}_{L\cup \{x\}}\\
\Tr{\be{x:t = e}e'}_L & \tconst{\Tr{t}}\; x = \Tr{e}_L; \Tr{e'}_L
\end{eqnarray*}
where the translation of types is $\Tr{b}=b$ and 
$\Tr{\tref{t}} = \tptr{\Tr{t}}$. The constant $\eref$ outside a
let-binding can be translated as |alloca| or |malloc|.
Compared to the extant translation, Fig.~\ref{f:offshoring-extant}, 
there are no longer any
restrictions on types. Adding constants like |incr| is easy.

The running example
\[
\be{x=\eref 0}x:={!x}+1
\]
now looks like
\[
\I\, x=0;\; \&x\leftarrow *\&x+1
\]
or, after rendering in C:
\[
\texttt{int x = 0;  x = x + 1;}
\]
The (extended)  problematic example
\[ 
\be{x=\eref 0}\be{y=x}y:=41;x:={!x}+1
\]
translates to CoreCE as
\[
\I\, x=0;\; \tconst{\tptr\I}\, y=\&x;\; y\leftarrow 41;\; \&x\leftarrow *\&x+1
\]

The new translation indeed gives idiomatic C code, which is easier to
inspect and build confidence. Unlike the translation of
\S\ref{s:no-mut-var}, only one CoreCE variable is allocated per ICaml
variable, with no extra pointer variables.  The extended calculus
CoreCE, and the current translation, which tracks mutability, stress
the fact that although all variables in C are mutable by default, some
are actually not mutated. The latter correspond to ICaml variables.
Actually mutable variables of CoreCE correspond to ICaml variables introduced
by the bindings
of a particular shape: $\be{x:\tref{t} = \eref e}e'$, which evoke
\textsf{letref} of the original ML.

With full details and formality the translation is presented in the
accompanying code \url{refcalculi.ml}. As mentioned earlier, the code
also states the denotational semantics $\Den{\cdot}_{ICaml}$ and
$\Den{\cdot}_{CoreCE}$, as compositional mappings from ICaml or CoreCE,
resp., to the common metalanguage, which is OCaml. (One could also use
Coq with a State monad.)
The translation from ICaml to CoreCE is then coded as a functor.
Since the (tagless-final) embeddings of ICaml and CoreCE into OCaml are
intrinsically typed, the fact that the translation functor is well-typed in
OCaml implies the translation is type-preserving. The meaning
preservation is expressed by the theorem that for each ICaml
expression $e$, $\Den{e}_{ICaml} = \Den{ \Tr{e} }_{CoreCE}$. To show it, we
have to check, manually at present, that the theorem holds for each
expression form of ICaml, and then appeal to compositionality of the
semantics.

\bigskip
In conclusion, we have learned that C variables are quite subtle: one
may access a mutable variable via its name or a pointer to it;
however, names and pointers are emphatically distinct.

The new offshoring translation, \S\ref{s:final-translation}, has been
implemented in BER MetaOCaml.

\paragraph{Acknowledgments}
We thank anonymous reviewers for many, helpful comments and
suggestions.
This work was partially supported by JSPS KAKENHI Grants Numbers
17K12662, 18H03218, 21K11821 and 22H03563.

\bibliographystyle{plainnat}
\bibliography{../metafx.bib}
\end{document}